\documentstyle[12pt]{article}

\newcommand{\be}{\begin{equation}}
\newcommand{\ee}{\end{equation}}

\newcommand{\ba}{\begin{eqnarray}}
\newcommand{\ea}{\end{eqnarray}}

\newcommand{\la}{\lambda}

\newcommand{\tr}{\rm tr}

\begin{document}

\hoffset=-.4truein\voffset=-0.5truein
\setlength{\textheight}{8.5 in}

\begin{titlepage}
\begin{center}
\hfill \\
\hfill {LPTENS-02/39}
\vskip 0.6 in
{\large An extension of the HarishChandra-Itzykson-Zuber integral}
\vskip .6 in
       \begin{center}
{\bf E. Br\'ezin$^{a)}$}{\it and} {\bf S. Hikami$^{b)}$}
\end{center}
\vskip 5mm
\begin{center}
{$^{a)}$ Laboratoire de Physique
Th\'eorique, Ecole Normale Sup\'erieure}\\ {24 rue Lhomond 75231, Paris
Cedex
05, France{\footnote{
 Unit\'e Mixte de Recherche 8549 du Centre National de la
Recherche
Scientifique et de l'\'Ecole Normale Sup\'erieure.
 } }}\\
 {$^{b)}$ Department of Basic Sciences,
}\\ {University of Tokyo,
Meguro-ku, Komaba, Tokyo 153, Japan}\\

\end{center}
\vskip 0.5 cm
{\bf Abstract}
 \end{center}
 \vskip 14.5pt

  The HarishChandra-Itzykson-Zuber integral over the unitary
  group $U(k)$ ($\beta=2$) is present in numerous problems involving
  Hermitian
random matrices. It is  well known that the result is
semi-classically exact.
  This simple result does not extend to other symmetry groups, such as the
symplectic or orthogonal groups. In this article the analysis of this
integral
  is extended first to the symplectic group $Sp(k)$ ($\beta$=4). There the
semi-classical approximation  has to be corrected by a WKB expansion. It
turns
out that this expansion stops after a finite number of terms ; in other
words
the WKB approximation is corrected by a polynomial in the appropriate
variables. The analysis is based upon
  new solutions to  the heat kernel differential equation.
   We have also investigated
   arbitrary values of  the parameter $\beta$, which characterizes the
   symmetry
group.
  Closed formulae are derived
  for arbitrary $\beta$ and $k=3$, and also for large $\beta$  and arbitrary
$k$.

\end{titlepage}
\setlength{\baselineskip}{1.5\baselineskip}
\section{  Introduction }

The HarishChandra and Itzykson-Zuber (HIZ) integral
\cite{Harish-Chandra,Itzykson-Zuber,Duistermaat}for the unitary group
$U_k$
is defined by
\be\label{gint}I = \int dU \exp{i \tr U^{\dagger} X U \Lambda}
\ee
where $X$ and $\Lambda$ are diagonal $k \times k$ Hermitian matrices.
It is well known that this integral is given, up to some normalization of the
measure, by
\be\label{HIZ1}
I = \frac{\det_{1\leq i,j\leq k}\exp(ix_i \lambda_j)}{\Delta ( X) \Delta(
\Lambda )}\ee where $\Delta(X) $ and $\Delta
(\Lambda)$ are the Vandermonde determinants of the eigenvalues of $X$ and
$\Lambda$, ($\displaystyle \Delta(X) = \prod_{i<j} (x_i - x_j)$). This HIZ
formula
may be easily derived by considering the Laplacian operator
\cite{Itzykson-Zuber} \cite{Brezin}

\be
L =  - \frac{\partial^2}{\partial X_{ij}^2}.
\ee

Its eigenfunctions are plane waves
\be L e^{ i \tr \Lambda X} = (\tr \Lambda ^2) e^{ i \tr \Lambda X}.
\ee One can construct a unitary invariant eigenfunction of $L$, for the
same eigenfunction
$\tr \Lambda^2$, by the superposition
\be I = \int dU e^{i\tr \Lambda U X U^{\dagger}},
\ee
 which is nothing but the HIZ integral. The integral $I$ beeing unitary
invariant,  is a function of the $k$ eigenvalues $ x_i$
of $X$. The one dimensional quantum mechanics of the eigenvalues, for
unitary invariant eigenstates, is then easily solved in terms of free
fermions, leading to the unitary HIZ formula.

 The
same considerations hold  for the three ensembles
$\beta= 1,2$ and 4, corresponding to the orthogonal, unitary  and
symplectic ensembles, with
\be
I = \int e^{i \tr \Lambda g X g^{-1}} dg.
\ee
The Laplacian acting on invariant states, may be expressed in terms of a
differential operator on the eigenvalues $x_i$ :
\be \label {7}
[ \sum_{i=1}^k \frac{\partial^2}{\partial x_i^2} + \beta \sum_{i=1,(i\ne
j)}^k
\frac{1}{x_i - x_j} \frac{\partial}{\partial x_i}] I =- \epsilon I\ ,
\ee
with the eigenvalue  $\epsilon$
\be
\epsilon =  \sum_{i=1}^k \lambda_i^2
\ee
Note that the integral $I$ is manifestly symmetric under interchange of
the matrices $\Lambda$ and $X$, but the procedure is dissymetric. The
solutions will of course   restore this  property, which is not obvious if
one
considers the equation  (\ref{7}) alone.

The
$x$-dependent eigenfunctions of this Schr\"{o}dinger operator have a scalar
product given by the measure
\be \langle \varphi_1\vert \varphi_2\rangle = \int dx_1\cdots dx_k
\vert\Delta(x_1\cdots x_k)\vert^{\beta} \ \varphi_1^{*}(x_1\cdots x_k)
\varphi_2(x_1\cdots x_k)
\ee
The measure becomes trivial if one
multiplies the wave function by
$\vert\Delta\vert^{\beta/2}$ . Thus  if, for some given ordering of the
$x_i$'s,
one changes
$I(x)$ to
\be \psi(x_1\cdots x_k) =
\vert\Delta(x_1\cdots x_k)\vert^{\beta/2} I(x_1\cdots x_k),  \ee one
obtains the
 Hamiltonian,
\be\label{r2}
[ \sum_{i=1}^k \frac{\partial^2}{\partial x_i^2} - \beta (\frac{\beta}{2}
 -1) \sum_{i<j}
\frac{1}{(x_i - x_j)^2}]\psi = -\epsilon \psi.
\ee
The relation between matrix quantum mechanics and many-body problems
with $1/r^2$ pair potentials was already present in \cite {BIPZ} and
it has also be used for the study of Selberg integrals by Forrester
\cite {Forrester}. This Schr\"{o}dinger equation is  a simple
Calogero-Moser model.

Although the solutions
of this many-body problem are known, they are of little use for our purpose,
since it is already quite involved to recover from the formal expression of
the
solution for arbitrary $\beta$,  the simple unitary result of HIZ and it
looks
formidable to extend it to other values of
$\beta$ \cite{Mahoux}.  For
$\beta =2$, the solution is again given by plane waves in the
$x_i$ and (taking into account the symmetry under permutations of $I$),
one obtains the HIZ formula.  For $\beta = 4$, the problem is much less
 trivial, but it leads to  explicit solutions for finite values of $k$.
Indeed the problem turns out to possess simple solutions of the form of
symmetrized sums of plane waves multiplied by polynomials in the variables
$1/(x_i-x_j)(\lambda_i-\lambda_j)$, providing therefore a complete
explicit solution for the $Sp(k)$ group integration.

For  general $k$, the solution of (\ref{r2}) is of the form
\be
  \psi_0 = e^{i (\lambda_1 x_1 + \cdots + \lambda_k x_k)} \chi
\ee
where $\chi$ satisfies
\be \label{diff}
[ \sum_{i=1}^k \frac{\partial^2}{\partial x_i^2} + 2 i \sum_{i=1}^k
\lambda_i
\frac{\partial}{\partial x_i} - y\sum_{i<j} \frac{1}{(x_i - x_j)^2}] \chi = 0
\ee
in which
\be \label{y} y = \beta (\frac{\beta}{2} -
1).\ee

For k=2, it is immediate to verify that there is a solution, function of
 the
signle variable $\tau = (\la_1-\la_2)(x_1-x_2)$, whose asymptotic
expansion
for large $\tau$ is
\ba \chi &=& 1 + \sum_{n=1}^{\infty} (-i/\tau)^n
\prod_1^{n}(p-1-\frac{y}{2p})\nonumber \\&=& I_{\nu}(z) \sqrt{2\pi z}
e^{-z}
\ea
in which the Bessel function $I_{\nu}(z)$ is defined with $\displaystyle
\nu^2 =
(y+1/2)/4$ and $z= i\tau/4$.

 When
$\beta =4$, the operator
$\displaystyle
\sum_{i=1}^k
\frac{\partial^2}
{\partial x_i^2}  - \sum_{i<j} \frac{4}{(x_i - x_j)^2}$ annihilates
the function $\Delta^{-1}(x_1,\cdots, x_k)$ . Consequently the solution
of (\ref{diff}) may be written
\be\label{f} \chi (x_1\cdots x_k) = \frac { f (x_1\cdots x_k) }{
\Delta (x_1\cdots x_k)} \ee
in which $f(x_1 \cdots x_k)$ is a polynomial of degree $k(k-1)/2$ in the
$x_a$'s. Defining
\be\label{tau} \tau_{ij} = (\la_i-\la_j)(x_i-x_j) \ee
one finds for $k=3$

  \ba\label{I3}
           \chi =&&
           1 + 2 i (\frac{1}{\tau_{12}}
           + \frac{1}{\tau_{23}}
           + \frac{1} {\tau_{31}})
            -4( \frac{1}{\tau_{12}\tau_{23}} +
\frac{1}{\tau_{23}\tau_{31}}
            + \frac{1}{\tau_{31}\tau_{12}} )\nonumber\\
            && - {12i}
\frac{1}{\tau_{12}\tau_{23}\tau_{31}} \ .
\ea

Remaining still with  $\beta = 4$, one may factor out the Vandermonde
determinant from
$\chi$ as (\ref{f}), and  one
obtains a differential equation for $f$,
\be\label{diffF}
\sum_{i=1}^k \frac{\partial^2}{\partial x_i^2} f +
2 (\frac{\beta}{2} - 1)
\sum_{i=1}^k (\frac{\partial f}{\partial x_i} + i \lambda_i f)
 (\Delta \frac{\partial}{\partial x_i} \frac{1}{\Delta})
 + 2 i \sum_{i=1}^k \lambda_i \frac{\partial f}{\partial x_i} = 0
 \ee
The solution of this equation is obtained by expanding in
 powers of the  $\lambda_i$'s,
but the expansion ends with the highest degree, the Vandermonde product of
degree six
$\Delta(\la_1
\cdots\la_4)$. Using the  notation of (\ref{tau}), $\tau_{ij} = (x_i -
x_j) (\lambda_i - \lambda_j)$,
 we obtain
\be\label{ff}
f = 1 - \frac{i}{4} ( \tau_{12}  + \tau_{13} + \tau_{14} + \tau_{23} +
\tau_{24}
+ \tau_{34}) + O(\tau^2)
\ee
as shown in the appendix of \cite{BHa}.
In the next section, we consider the symplectic case ($\beta=4$)
for  larger values of $k$, and we will clarify the meaning
of the coefficients which appear in the expansion of $\chi$.

\section{Decomposition into complete graphs in the symplectic
case}
Let us discuss now the asymptotic expansion of $\chi$ for  $ \beta=4$,
i.e. for the symplectic group integration of the HIZ integral. Such
integrals
appear in various problems ;
for instance one encounters it  in the calculation of the average of
products of  characteristic polynomials of random matrices  in the Gaussian
orthogonal ensemble
$<\prod_{i=1}^k {\rm det} (\lambda_i - X)>$. Indeed we have shown in a
previous
work \cite{BHa} that   this average may be expressed as
an integral over a
$k\times k$ quaternion matrix ; this representation exhibits a duality
between
the GOE average and a symplectic HIZ integral. Thus this requires
to compute the HIZ integral for
$\beta=4$,  as discussed in \cite{BHa} , in which the expressions were
given up to
$k=4$. We extend here the
construction of the polynomial solutions  to  arbitrary values of  $k$ .

 As discussed before, the expansion of the function $\chi$ in inverse
powers of the variables $1/\tau_{ij}$, defined in (\ref{tau}), terminates
with a term  proportional to
$\displaystyle
\prod_{i<j}\frac{1}{\tau_{ij}}$, of degree $k(k-1)/2$.
The monomials in the variables $1/\tau_{ij}$ may be represented
graphically : one marks
$k$ points, and a line between the points i and j will represent a factor
$\frac{1}{\tau_{ij}}$ in a given monomial.
A {\it{complete}} graph is a graph in which all the pairs of points
are connected
by a line.
For instance, a triangle (k=3) and a tetrahedron (k=4) are complete
graphs.
The k-point complete graph  has  k(k-1)/2 lines.
A complete graph of k points receives a factor  $C_k$ in which
\be
C_k = \prod_{l=1}^k l!
\ee
This number is related to the integral,
\be
C_k = \frac{1}{(2 \pi)^{k/2}}
\int \prod_{i<j} ( z_i - z_j)^2 e^{- \frac{1}{2} \sum z_i^2} \prod_{i}^k
d z_i
\ee
The last term of highest degree in $\chi$
, is equal to
\be
i^k C_k \prod_{i<j}^k \frac{1}{\tau_{ij}}
\ee
All the  terms of the expansion of $\chi$ have  coefficients which
may be  decomposed into combinations of factors $C_n$ with $n\leq k$.

For instance ,
 the k=3 solution is
\be
\chi = 1 + i \frac{C_2
C_1}{C_0}
(v_1 + v_2 + v_3) + i^2 \frac{(C_2)^2}{C_1}(v_1 v_2 + v_2 v_3 +
v_1 v_3)
+ i^3 C_3 v_1 v_2 v_3\ee
where $\displaystyle v_3 = \frac{1}{\tau_{12}} $, $\displaystyle v_1 =
\frac{1}{\tau_{23}}$,and $\displaystyle v_2 = \frac{1}{\tau_{31}}$.
The graphs
of  the various terms
in $\chi$  may all be analyzed as  a decomposition into complete graphs. For
instance,
$v_1 v_2$ consists of two lines connecting the point number three to points
one and two. The lines (2-3) and (3-1) are complete graphs of two points,
hence the factor $C_2^2$, with the point three counted twice, thus a
division by
$(C_1)$, equal to one. (One has to multiply further the result
 by a power  of
$i$ raised to the number of lines).

 When $k=4$, by this rule of  decomposition of the graphs into
complete
graphs, we find the coefficients of all the terms in the form
\ba
\chi &=& 1 + i \frac{C_2 C_1}{C_0}
\sum_{i=1}^6 v_i \nonumber\\
&&
+ i^2 \frac{(C_2)^2}{C_1}(v_1 v_2 + v_2 v_3 + v_1 v_3 + v_1 v_4 +v_1
v_5 +v_2
v_5 + v_2 v_6 + v_3 v_6 \nonumber\\
&& + v_3 v_4 + v_4 v_5 + v_4 v_6 + v_5 v_6)\nonumber\\
&& +
i^2 (C_2)^2
(v_1 v_6 + v_2 v_4 + v_3 v_5)\nonumber\\
&& + i^3 \frac{C_3
C_1}{C_0} (v_1 v_2 v_3 + v_1 v_4 v_5 + v_2 v_5 v_6 + v_3 v_4
v_6)\nonumber\\
& & + i^3 \frac{(C_2)^3}{(C_1)^2}(v_1 v_3 v_5
+ v_1
v_3 v_6 + v_2 v_3 v_4 + v_3 v_4 v_5 + v_3 v_5 v_6 + v_1 v_5 v_6 + v_2
v_4 v_5
\nonumber\\
&&+ v_2 v_3 v_5
+ v_1 v_2 v_4 + v_1 v_4 v_6 +
v_1 v_2
v_6 + v_2 v_4 v_6)\nonumber\\
&& + i^3 \frac{(C_2)^3}{C_1}
(v_1 v_2
v_5 + v_1 v_3 v_4 + v_2 v_3 v_6 + v_4 v_5 v_6)\nonumber\\
& &
+ i^4
\frac{(C_2)^4 C_0}{(C_1)^4}(v_1 v_2 v_4 v_6 + v_1 v_3 v_5 v_6 + v_2 v_3 v_4
v_5)\nonumber\\
&& + i^4 \frac{C_3 C_2}{C_1}
(v_2 v_4 v_5 v_6 +
v_2 v_3 v_5 v_6 + v_1 v_2 v_5 v_6 + v_3 v_4 v_5 v_6 + v_1 v_3 v_4
v_6 + v_2
v_3 v_4 v_6\nonumber\\
&& + v_1 v_2 v_3 v_5 + v_1 v_2 v_3
v_6 + v_1
v_2 v_3 v_4 + v_1 v_3 v_4 v_5 + v_1 v_2 v_4 v_5 + v_1 v_4 v_5
v_6)\nonumber\\
&& + i^5 \frac{(C_3)^2}{C_2}( v_2 v_3 v_4
v_5 v_6
+ v_1 v_3 v_4 v_5 v_6 + v_1 v_2 v_4 v_5 v_6 + v_1 v_2 v_3 v_5 v_6 \nonumber
\\
&& +
v_1 v_2
v_3 v_4 v_6 + v_1 v_2 v_3 v_4 v_5)\nonumber\\&& + i^6 C_4
v_1 v_2
v_3 v_4 v_5 v_6
\ea
where we have
used  $\displaystyle v_1=
1/\tau_{12},v_2=
1/\tau_{23},v_3=1/\tau_{13},v_4 = 1/\tau_{14},v_5=1/\tau_{24},v_6=
1/\tau_{34}$.
Again the graph of the last term $\displaystyle v_1 v_2 v_3 v_4 v_5
v_6$ is a
complete graph
of 4-points.
Substituting in this formula the values $C_n = \prod_{j=1}^n j!$, one
recovers precisely the $k=4$  result,  given in an earlier
publication\cite{BHa,Guhr}.

The decomposition of graphs into complete
components has certain general structures for arbitrary values of $k$. The
term of highest degree is
$ i^{k(k-1)/2} C_k$ ; the next one,  obtained by deleting one
line, has a
decomposition given by \be\label{factor1}
\frac{(C_{k-1})^2}{C_{k-2}}\ee
For k=5, this decomposition
means that the graph is made of two
tetrahedra sharing a triangular face ; hence the division by $C_3$ to
avoid double counting of this face. This explains the
above factor  (\ref{factor1}). In the case k=4, this factor
represents two
triangles
connected by one edge, and it reads
$\displaystyle \frac{(C_3)^2}{C_2}= 72$.

When two lines are eliminated from the
maximum
complete graph $C_k$, we have two type graphs.

(i) The first case is the elimination
of  two
lines emerging from the same point. In this case the  decomposition  is
$\displaystyle C_{k-1}C_{k-2}/C_{k-3}$. For k=4, it is a triangle and one
line connected at the
vertex. For k=5, it is a tetrahedron connected to a
triangle through
one edge.

(ii) The second type  consists of deleting two non consecutive
lines.  In this case, we have the
decomposition of
$\displaystyle (C_{k-2})^4 C_{k-4}/(C_{k-3})^4$. In the
case k=4,
it corresponds just to a square, four line connecting the four vertices of a
square. For k=5, it is a graph of four triangles connected to each other
by four edges.

If we
divide this factor $\displaystyle (C_{k-2})^4 C_{k-4}/(C_{k-3})^4$
by $\displaystyle C_k$, as appeared in the expression for
$f$ in (\ref{ff}), it
becomes
$\displaystyle \frac{1}{18}, \frac{3}{80}, \frac{2}{75}$
for k= 4,5 and 6, respectively. These values have been found
by a direct evaluation of the solution of the differential equation
(\ref{diffF}). Thus we have, for
$\beta=4$
and for arbitrary values of k, an expression for $\chi$, whose
coefficients
are given by the decomposition of the complete graphs $C_n$.

The
case
$\beta=4$ is thus easy to analyze for two reasons. First there are no
"multiple" lines in a graph, meaning that every monomial in $\chi$ is of
degree at most one in any variable $1/\tau_{ij}$. Then the factor
$C_n = \prod j!$ for   complete graphs has a simple structure. However, for
$\beta\neq 4$,  the number of lines between two  points may be arbitrarily
large. Then the contribution $\displaystyle C_n$ of complete graphs
has a more complicated expression.
For values of   $\beta$ of the form
\be \label{m}\beta = 2(m+1)\ee
 a complete graph between two points
consists of  $m$ lines  joining these two points and
\be C_2 =
\frac{y}{2}( -1 + \frac{y}{4}) (-2 + \frac{y}{6}) \cdots (-(m -1) +
\frac{y}{2m})\ee
in which  y has be defined in (\ref{y}).
The evaluation of $C_3$ is non trivial and we
will
discuss it in the following section.
For $\beta$'s of the form (\ref{m})
the previous graphical analysis of decomposition into complete subgraphs,
may
be extended to
arbitrary k. However, even then,  more complex rules for
non-complete graphs are required. To understand this complexity,
we consider the $k=3$ case for arbitrary $\beta$
in the next section.

\section{k=3 for arbitrary $\beta$}

Even for the simple $k=3$ case, the HIZ integral  is not known
explicitly in the form of a WKB expansion. An analysis in terms of the
solutions of the  Calogero three-body problem has been performed in
\cite{Mahoux}.
 However, its outcome is far from our purpose
since it is already quite involved to recover from there the unitary HIZ
formula, and  one does not see that for values  of $\beta$ such as four, the
WKB expansion is a polynomial.
Therefore we start again with the partial derivative equation (\ref{diff})
for $k=3$,  and expand the solution as
\be\label{chi}
\chi
= \sum_{n,m,r=0} i^{n + m + r}\frac{C_{n,m,r}}{\tau_{12}^n
\tau_{23}^m
\tau_{31}^r}.\ee
Note that it is not at all obvious  that the equation (\ref{diff}) admits
such a solution. Indeed, assuming that the solution is only a function of
the $\tau$'s, one finds easily that
\be[ 2 i \sum_{i=1}^3
\lambda_i
\frac{\partial}{\partial x_i} - \sum_{i<j} \frac{
y}{(x_i - x_j)^2}] \chi =
-(\la_1-\la_2)^2\tau_{12}^2[2 i\frac{\partial}{\partial\tau_{12}} +y]
\chi+\rm{perm} .\ee
 However
the term involving the second derivative does not share this structure since
\ba \sum_1^3 \frac{\partial^2 \chi}{\partial x_a^2}
=&&2(\la_1-\la_2)^2[\tau_{12}^4\frac{\partial^2\chi}{\partial
\tau_{12}^2}
+\tau_{12}^3\frac{\partial}{\partial \tau_{12}}]\nonumber \\
&&-2(\la_1-\la_2)(\la_3-\la_1)
\tau_{12}^2\tau_{31}^2 \frac{\partial^2\chi}{\partial \tau_{12}\partial
\tau_{31}} +\rm{perm}  \ea
The last term, with cross-products in the $\la_a$'s, is a priori different.
However there are, order by order, a set of identities which allow one to
cast
those cross-products into  sums of squares. Those identities
follow from $(x_1-x_2)+(x_2-x_3)+(x_3-x_1) = 0$, leading to
\be (\la_1-\la_2) \tau_{23}\tau_{31} + \rm{perm} = 0.\ee
From there one derives order by order the identities which are needed to
cast the cross-products into squares. Let us quote the simplest
\be\label{id0} (\la_1-\la_2)(\la_3-\la_1) \tau_{23}^2+ \rm{perm} =
(\la_1-\la_2)^2\tau_{31}\tau_{23}+\rm{perm}.\ee
Order by order those identities allow us to satisfy the conditions on the
expansion of $\chi$, which make it  a function of the $\tau$ variables .
 One finds easily  that
$C_{n00}$,  which is  associated to the diagram consisting of $n$ lines
between two points, is given by  :
\be C_{n00} =
\prod_{l=1}^n ( - (l -
1) + \frac{y}{2 l})
\ee
The generating function of those coefficients  is the
modified Bessel function, solution of the k=2 problem.
Note that
$C_{n,m,r}$ is symmetric under permutations of n,m and r ;
for instance  $C_{m,0,0} = C_{0,m,0}$.

 When r=0, one finds that $C_{n,m,0}$ has  a
decomposition
into a product of two factors,
\be
C_{n,m,0} = C_{n,0,0}
C_{m,0,0} . \ee

For the full  HIZ
integral,
triangle graphs, in which the three integers $(n,m,r)$ of $C_{n,m,r}$ are
non-zero, play an essential role.

From the differential equation (\ref{diff}), one can derive a recursion
equation for the
$C_{n,m,r}$ . However a direct expansion of the solution of
 is extremely tedious and complicated at increasing orders, since one has to
use more and more identities of the type (\ref{id0}), but of higher degree.
One needs triagle identities. The expansion for the
differential equation  (\ref{diffF}) has a simpler structure ; it
gives a
relation between $C_{n,m,r}$, with $n+m+r = P+1$, in terms of $C_{n,m,r}$
with $n+m+r= P$ ; this relation is valid for a fixed given value of $m$.
If
one writes
$C_{n,m,r}$ as a  linear combination of $C_{n^{\prime},m,r^{\prime}}$,
where $r^{\prime} < r$, and $n^{\prime}+r^{\prime} = n+r-1$, one obtains

\be
C_{n,m,1} = C_{n,m,0}(n m + \frac{y}{2})\ee
\be
C_{n,m,2} =
 C_{n,m,1}
(\frac{m(n-1)}{2} - 1 + \frac{y}{4}) + C_{n+1,m,0} \frac{m}{2}(n +
1)
\ee
\ba C_{n,m,3} &= & C_{n,m,2}(\frac{m}{3}(n - 2)
- 2 +
\frac{y}{6}) + C_{n+1,m,1}
\frac{mn}{3}\nonumber\\
&& + C_{n+2,m,0} \frac{m}{3}(n + 2)
\ea
For general $r$ ($r>3$),
we
have
\ba\label{recursion}
C_{n,m,r}  && =
C_{n,m,r-1}(\frac{m
(n - r + 1)}{r} - ( r - 1) +
\frac{y}{2 r})\nonumber\\
&& + C_{n,m,r-2} \frac{m}{r}(n - r + 3) +
C_{n,m,r-3}\frac{m}{r}(n - r +
5)\nonumber\\
&& + \cdots
\nonumber\\
&&+ C_{n+r-2,m,1}\frac{m}{r}(n + r - 3) +
C_{n+r-1,m,0}\frac{m}{r}(n +
r - 1).\ea
The solution of these recursion equations guarantees
automatically the symmetry under  exchange between $n,m$ and
$r$ :
$C_{n,m,r} = C_{n,r,m} = C_{m,n,r} = C_{m,r,n} = C_{r,n,m}
= C_{r,m,n}
$.

From these recursion formulae, one may determine iteratively
all the coefficients
$C_{n,m,r}$. The resulting expressions are complicated  for general
$y$.
For instance,  even at low order, one finds
\be C_{2,2,2} =  (\frac{y}{2})^2(-1 + \frac{y}{4})^2 ( - 6 +
\frac{3}{2}
y + \frac{y^2}{8})
\ee
\be
C_{3,3,3} =  (\frac{y}{2})^2 (- 1 +
\frac{y}{4})^2 (-2 + \frac{y}{6})^2 (-51 - \frac{59}{8}y +
\frac{37}{48}y^2 +
\frac{1}{48}y^3)
\ee
The solution for  $\beta = 4$ and  $k=3$, may be written in the compact
form
\be
\chi = e^{\phi}
\ee
\be \phi =
2 i
(\frac{1}{\tau_{12}} + \frac{1}{\tau_{23}} + \frac{1}{\tau_{13}})
 - 4 i
\frac{1}{\tau_{12}\tau_{23}\tau_{13}}\ee
in which it is understood that one expands it in powers of $1/\tau$,
dropping
any term involving  powers of  $1/\tau_{ij}$ greater than one.
This corresponds to the absence of multiple bonds between two points in the
graphical representation for  $\beta=4$. This result agrees with (\ref{I3}).
This certainly suggests that $\phi$ ought to be written in terms of
Grassmann
variables. In the next section, we
consider such expressions for general $y$ in an
expansion in $1/y$.
This also provides consistency checks for the recursion equation
(\ref{recursion}).
\section{Large y expansion}
One can obtain an expansion in powers of $1/y$ if one returns to the
differential equation (\ref{diff}), taking  now as variables
\be  u_a = x_a/y.\ee
The function $\chi(u_1\cdots u_k)$ is solution of
\be \label {diffy} [\frac{1}{y}  \sum_{a=1}^k \frac{\partial^2}{\partial
u_a^2}
+ 2 i \sum_{a=1}^k
\lambda_a
\frac{\partial}{\partial u_a} - \sum_{a<b} \frac{1}{(u_a - u_b)^2}] \chi =
0.\ee
The leading term $\chi_0$ is thus solution of
\be\label{41} [2 i \sum_{a=1}^k
\lambda_a
\frac{\partial}{\partial u_a} - \sum_{a<b} \frac{1}{(u_a - u_b)^2}] \chi_0
=0.\ee

We look for a solution which is a symmetric function of the variables
\be \label{x} v_{a,b}  = \frac{ i}{(\lambda_a-\la_b)(u_a-u_b)} \ ,\ a<b
\ee
 which thus satisfies
\be \label {43}  \sum_{a=1}^k \la_a \frac{\partial \chi_0}{\partial u_a}
=
i\sum _{1\leq a<b\leq k} (\la_a-\la_b)^2 v_{ab}^2 \frac{\partial
\chi_0}{\partial v_{ab}}.\ee
The equation (\ref{41}) takes then the simple form
\be\sum _{1\leq a<b\leq k} (\la_a-\la_b)^2 v_{ab}^2 [2\frac{\partial
}{\partial v_{ab}} -1]\chi_0 = 0.\ee
There is a trivial symmetric solution which satisfies the $k(k-1)/2$
equations
 \be [2\frac{\partial
}{\partial v_{ab}} -1]\chi_0 =0, \ee
namely
\be \label{zero}  \chi_0 = \exp{( \frac{1}{2}\sum_{a<b} v_{ab})} =
\exp{ (\frac{iy}{2}\sum_{a<b} \frac{1}{(\la_a-\la_b)(x_a-x_b)})}.\ee
One thus writes
\ba  \label{g} \chi &=& e^\phi \nonumber \\
\phi &=& \phi_0 + \frac{1}{y} \phi_1 + \frac{1}{y^2}\phi_2 +\cdots ,\ea
with
\be \label{zero'}\phi_0 = \frac{1}{2}\sum_{a<b} v_{ab}. \ee
The next terms are solutions of
\be \label{phi1} \sum_1^k ((\frac{\partial \phi_0}{\partial u_a})^2 +
\frac{\partial^2
\phi_0}{\partial {u_a}^2}) +2i\sum_1^k \la_a \frac{\partial \phi_1}{\partial
u_a} = 0,\ee
\be \label{phi2} \sum_1^k (2\frac{\partial \phi_0}{\partial u_a}\frac{
\partial
\phi_1}{\partial u_a} +
\frac{\partial^2
\phi_1}{\partial {u_a}^2}) +2i\sum_1^k \la_a \frac{\partial \phi_2}{\partial
u_a} = 0.\ee
In order to make it clear that it is a non trivial property of these
equations
to have solutions which are symmetric polynomials in the variables
$v_{ab}$,
we shall examine these equations in some detail.

For simplicity of notations we shall limit ourselves to k=3, the
generalization
to arbitrary k beeing immediate. It is thus convenient to simplify
slightly
the
notations and write
\be v_{12} = V_3\ ,\ v_{23} = V_1\ ,\ v_{31} = V_2 \ee
Let us first consider (\ref{phi1}) ; since the function $\phi_1$ is a
symmetric
function of the  $V_i$'s one finds easily, as in (\ref{43}), that
\be \label{52}\sum_1^3 \la_a \frac{\partial \phi_1}{\partial
u_a} = i [(\la_2-\la_3)^2 V_1^2 \frac{\partial}{\partial V_1} +
\rm{perm.}]\phi_1\ee
The explicit solution (\ref{zero'}) allows one to write
\ba \sum_1^3 ((\frac{\partial \phi_0}{\partial u_a})^2
 &=& -\frac{1}{2} [ (\la_2-\la_3)^2 V_1^4+ \rm{perm.}] + \frac{1}{2}[
(\la_1-\la_2)(\la_3-\la_1) V_2^2 V_3^2 + \rm{perm.}]
\nonumber\\
\sum_1^3  \frac{\partial^2 \phi_0}{\partial u_a^2}
 &=& -2 [ (\la_2-\la_3)^2 V_1^3+ \rm{perm.}]\ea

The second term of the first line of (\ref{52}) does not involve a priori
$(\la_a-\la_b)^2$ ; however the variables $V_i$ are not independent since
they
are defined in terms of the differences  $(t_1-t_2),(t_2-t_3),(t_3-t_1)$
whose sum is zero. Thus
\be \label{id1} V_2V_3 (\la_1-\la_2)(\la_1-\la_3) +\rm{perm.} = 0\ee
and  one can
check without difficulty  that
\be \label {id2} V_2^2V_3^2 (\la_1-\la_2)(\la_1-\la_3) +\rm{perm.} =
-V_1V_2V_3[(\la_2-\la_3)^2 V_1+\rm{perm.}]. \ee
With the help of this identity the equation for $\phi_1$ takes the simple
form
\be (\la_2-\la_3)^2V_1^2[ \frac{\partial \phi_1}{\partial V_1} + \frac{1}{4}
V_1^2 + V_1 -\frac{1}{4} V_2V_3] + \rm{perm.} = 0 .\ee

Thanks to the identity (\ref{id2}) one can look for a solution of the three
equations
\be  \frac{\partial \phi_1}{\partial V_1} + \frac{1}{4}
V_1^2 + V_1 -\frac{1}{4} V_2 V_3 = 0 \ee
plus the other two deduced from the cyclic permutations. The system is
trivially
integrable and one finds
\be \phi_1 = \frac{1}{4} V_1V_2V_3 -\frac{1}{2}(V_1^2+V_2^2+V_3^2)
-\frac{1}{12}(V_1^3+V_2^3+V_3^3) .\ee
For arbitrary $k$ this generalizes to
\be \phi_1 = \frac{1}{4} \sum_{(abc)}v_{ab}v_{bc}v_{ca}
-\frac{1}{2}\sum_{a<b}v_{ab}^2 -\frac{1}{12}\sum_{a<b}v_{ab}^3
,\ee
in which  the sum (abc) runs over all distinct triplets of indices.

We have
exposed this simple calculation in some detail, in order to make it clear
that the success of the method  relies on two successive facts :
\begin{itemize}
\item Identities such as (\ref{id1}, \ref{id2}) which allows one to write
the
differential equation as three separate equations for $\partial
\phi/\partial
V_a$
\item The integrability conditions of this system of three equations  are
satisfied.
\end{itemize}
At higher orders the technique is identical. It requires new identities of
higher degree, such as
\ba \label {id3}&& V_2^2V_3^2(V_2^2+V_3^2) (\la_1-\la_2)(\la_1-\la_3)
+\rm{perm.}
\nonumber \\&& = -(\la_2-\la_3)^2V_1V_2V_3[ V_1(V_1^2+V_2^2+V_3^2) +
\frac{1}{2} V_1V_2V_3]+\rm{perm.}.
\ea
and many others ;   thereby one obtains again remarkably an integrable
system
of three equations. At order
$1/y^2$ the solution is
\ba \phi_2 = &&(V_1^3+V_2^3+V_3^3) + \frac{1}{2}
(V_1^4+V_2^4+V_3^4)-\frac{1}{2}V_1V_2V_3( V_1+V_2+V_3)\nonumber \\
&&+\frac{1}{20} (V_1^5+V_2^5+V_3^5)-\frac{1}{8}V_1V_2V_3 (V_1^2+V_2^2+V_3^2)
\ea

\section{ A cubic identity}

  Let us consider the case of $k=4$ for
  different values of $\beta$, assuming that it is an even integer. For the
tetrahedron (k=4),
  the coefficients of the expa nsion of $\chi$ may be written as
  \be\label{C}
    \chi = \sum_{n,m,r,p,q,l=0}^j i^{n+m+r+p+q+l}
    \frac{C_{n m r;p q l}}{\tau_{12}^n \tau_{13}^m
    \tau_{23}^r \tau_{24}^p \tau_{34}^q \tau_{14}^l}
  \ee
  where $j=\beta/2 -1$.
  By imposing to satisfy the differential equation  (\ref{diffF}), one
  should determine those coefficients. For instance, in the case $\beta=6
$ (or $y=12$),  one finds
  $\displaystyle C_{222;222} = 8\cdot 7 (6\cdot 5)^2
  (4\cdot 3)^3, C_{222;221}=(2\cdot 7)(6\cdot 5)^2
  (4\cdot 3)^3, C_{222;211} = 4 (6\cdot 5)^2 (4\cdot 3)^3,
  C_{222;220} = (C_{222;000})^2/C_{200;000}= (6\cdot 5)^2
  (4\cdot 3)^3, C_{222;000} = (6\cdot 5)(4\cdot 3)^2,
  C_{221;221} = 6 (6\cdot 5)^3 2^5,
  C_{nm2;pq1} = C_{nm2;pq0}(np + mq + \frac{y}{2}),
  C_{nmr;pq0} = C_{nmr;000}C_{pqr;000}/C_{r00;000},\\
  C_{222;pq2}=C_{222;pq1}( \frac{p}{2} +
  \frac{q}{2} - 1 + \frac{y}{4})$.
  (The coefficient $C_{nmr;000}$ is identical to the $C_{nmr}$ of the k=3
case
  (\ref{chi}). Whenever the graphs reduce  to triangles,
  one recovers the  coefficients of the  k=3 problem).

    However the above representation (\ref{C}) is in fact somewhat
ambiguous. Indeed there is an interesting cubic identity between the six
variables $$\tau_{ab}= (x_a-x_b)(\la_a-\la_b)\ ,1\leq a<b\leq 4 , $$ which
holds for every set of
$x_a$'s and $\la_a$'s, namely
\ba
&&\tau_{12}^2 \tau_{34} + \tau_{13}^2 \tau_{24} + \tau_{14}^2 \tau_{23}
+ \tau_{23}^2 \tau_{14} + \tau_{24}^2 \tau_{13} + \tau_{34}^2 \tau_{12}
\nonumber\\
&& - \tau_{12}\tau_{34}( \tau_{23} + \tau_{24} + \tau_{14} +
\tau_{13} ) -
\tau_{13}\tau_{24}( \tau_{14} + \tau_{12} + \tau_{23} + \tau_{34} )
\nonumber\\
&& - \tau_{14}\tau_{23}( \tau_{12} + \tau_{24} + \tau_{34} + \tau_{13} )
\nonumber\\
&& + \tau_{12}\tau_{24}\tau_{14} + \tau_{13}\tau_{14}\tau_{34} + \tau_{23}
\tau_{24}\tau_{34} + \tau_{12}\tau_{23}\tau_{13} = 0 .
\ea

Consequently, when one divides the l.h.s. of this  identity by the
Vandermonde squared $\Delta(x)^2 \Delta(\lambda)^2$, one generates  an
identity for the terms of the exapnsion of $\chi$  whose  coefficients are
$\displaystyle C_{221;220},C_{221;211},C_{122;121}$.
Thus there is an arbitrariness in  writing $\chi$ in the form of the expansion
(\ref{C}), in view of  this  cubic identity. For arbitrary $k$, there are
$(k-2)(k-3)/2$ similar cubic identities, and therefore only $(2k-3)$
independent $\tau_{ab}$.

For other values of $\beta$ such as $\beta=8$ for instance, by imposing
the differential equation  (\ref{diffF}),  one finds that the
values of $\displaystyle C_{332;331},C_{332;322},C_{323;322}$
are not uniquely determined, because again of
this same cubic identity. One may need to eliminate this arbitrariness
by choosing some arbitrary value for some coefficient, decide that is it zero
for instance.
This cubic identity carries  the problem over to higher orders. Indeed,
multiplying  this same cubic identity, by some symmetric function of the
$\tau$'s, such as
$\displaystyle \sum \tau_{ij}$,
one obtains an identity of order $\tau^4$, which again requires some
arbitrary choice in  the  expression for
$\chi$ at next order.

In summary, we have obtained  expressions for the
HarishChandra-Itzykson-Zuber integration for $k\times k$
matrices, for group integrals characterized by the parameter $\beta$. Our
investigation relies upon new solutions to the  heat kernel differential
equation. Other methods  such as supersymmetric quantum mechanics,
combinatorial group analysis, or  algebraic geometry   may be of help in
solving the numerous open questions concerning this problem.

{\bf Acknowledgements}

 It is a pleasure to thank Dr. Alfaro for  interesting remarks concerning
the relevance of supersymmetric quantum mechanics in this problem. S.H.
has benefited from  a Grant-in-Aid for Scientific Research (B) by JSPS.



\begin{thebibliography}{99}
\bibitem{Harish-Chandra}
Harish-Chandra, Proc. Nat. Acad. Sci. {\bf
42}, 252(1956).
\bibitem{Itzykson-Zuber} C. Itzykson and J.-B.
Zuber,
J. Math. Phys. {\bf 21}, 411 (1980).
\bibitem{Duistermaat} J. J.
Duistermaat and G. H. Heckman, Invent. Math.{\bf 69},
259 (1982).
\bibitem{BHa}
E. Br\'ezin and S. Hikami, Commun. Math. Phys. {\bf
223}, 363
(2001).
\bibitem{BHb} E. Br\'ezin and S. Hikami, Commun. Math.
Phys.
{\bf 214}, 111 (2000).
\bibitem {Guhr} T. Guhr and H. Kohler,
J. Math. Phys. {\bf 43}, 2707 (2002).
\bibitem{BIPZ} E. Br\'ezin, C. Itzykson, G.
Parisi and J.-B. Zuber, Comm. Math. Phys.{\bf 59}, 35 (1978).
\bibitem{Forrester} P.J. Forrester, Nucl. Phys. {\bf B388}, 671 (1992).
\bibitem{Brezin} E. Br\'ezin, "Two dimensional quantum
gravity
and random surfaces", p.37, edited by D. J. Gross T. Piran and S.
Weinberg,(1992),World Scientific, Singapore.
\bibitem{Mahoux} G. Mahoux, M. L. Mehta and J.-M. Normand,
     "Random Matrices and Their Applications",
     MSRI Publications, {\bf 40}, 301 (2001).
\end{thebibliography}
\end{document}